\documentclass{v16nufact}

\def\Journal#1#2#3#4{{#1} {\bf #2}, #3 (#4)}

\def\NIMA{{\em Nucl. Instrum. Methods} A}

\def\PLB{{\em Phys. Lett.}  B}
\def\PRL{\em Phys. Rev. Lett.}
\def\PRD{{\em Phys. Rev.} D}

\def\RSI{\em Rev. Sci. Instrum.}
\def\PhysProc{\em Physics Procedia}

\def\be{\begin{equation}}
\def\ee{\end{equation}}
\def\bea{\begin{eqnarray}}
\def\eea{\end{eqnarray}}

\usepackage{xspace}

\newcommand{\DD}{DAE$\delta$ALUS\xspace}
\newcommand{\htp}{\ensuremath{\mathrm{H}_2^+}\xspace}
\newcommand{\nuebar}{\ensuremath{\bar{\nu}_e}\xspace}

\newcommand{\numubar}{\ensuremath{\bar{\nu}_{\mu}}\xspace}

\newcommand{\hminus}{\ensuremath{\mathrm{H}^-}\xspace}
\def\etal{$\it {et} \it{al}$}
\usepackage{wrapfig}
\usepackage[nodisplayskipstretch]{setspace}
\usepackage[font=small,skip=0pt]{caption}

\newcommand{\Photo}{}

\begin{document}
\vspace*{4cm}
\title{DEVELOPMENTS FOR THE ISODAR@KAMLAND AND \DD DECAY-AT-REST NEUTRINO EXPERIMENTS}

\author{ JOSE R. ALONSO FOR THE ISODAR COLLABORATION }

\address{Massachusetts Institute of Technology, 77 Massachusetts Avenue,\\
Cambridge, MA, 02139, USA}

\maketitle\abstracts{
Configurations of the IsoDAR and \DD decay-at-rest neutrino experiments are described. Injector and cyclotron developments aimed at substantial increases in beam current are discussed.  The IsoDAR layout and target are described, and this experiment is compared to other programs searching for sterile neutrinos.}

\section{Introduction}

  \begin{wrapfigure}{R}{0.45\textwidth}
\vspace*{-5mm}
  \includegraphics[width=0.45\textwidth]{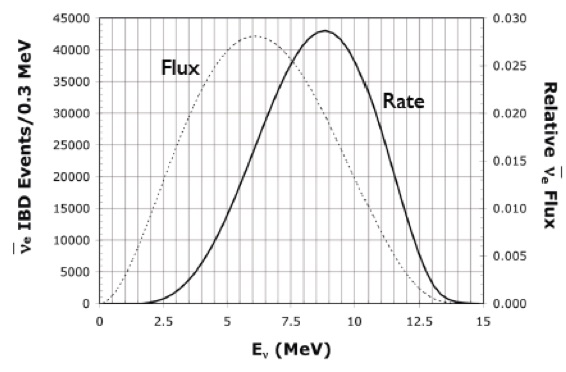}
\vspace*{-8mm}
  \caption{ $^8$Li neutrino spectrum.  Dashed = actual spectrum, 
  Solid = detector response for IBD events
    \label{Li8nuspectrum}}
    \vspace*{-6mm}
  \end{wrapfigure}

Decay-At-Rest (DAR) experiments offer attractive features for neutrino physics studies.\cite{DAR}    
  We discuss two particular regimes where the characteristics of the source are 
determined by the nature of the weak-interaction decay producing the neutrino,
 and are not affected by kinematics or characteristics of higher-energy production 
 mechanisms.  The beta decay case is manifested in the IsoDAR experiment; a
  sterile-neutrino search where a 60 MeV proton beam is used to produce the 
  parent isotope, $^8$Li.  
  The product nucleus is stationary when it decays, the 
  neutrino spectrum is shown in Figure \ref{Li8nuspectrum}.  It has a high endpoint energy, 
  over 13 MeV, 
  and a mean energy of 6.5 MeV, both substantially higher than backgrounds from other decays, 
  and in an area easily accessible for detection by Inverse Beta Decay (IBD) in a hydrogen-containing 
  neutrino detector.  
    
   \begin{wrapfigure}{l}{0.4\textwidth}
  \begin{center}
  \vspace*{-7mm}
  \includegraphics[width=0.4\textwidth]{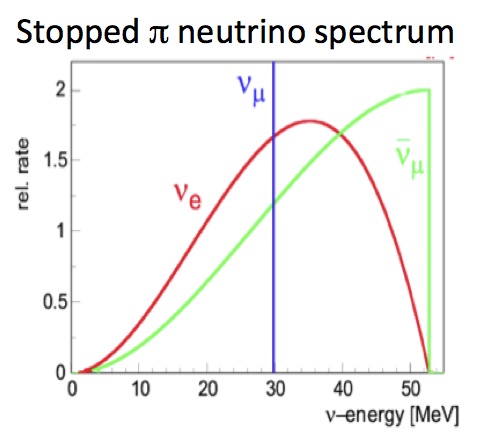}
  \end{center}
  \vspace*{-6mm}
  \caption{ Neutrino spectrum from stopped $\pi^+$.  Note absence of \nuebar.
    \label{pionspectrum}}
    \vspace*{-5mm}
  \end{wrapfigure}

  
  In the regime where pions are produced at low energy (with $\le 800$ MeV protons), pions can stop in the target before decaying.  This is the case for \DD, a sensitive CP violation measurement.  As the nuclear capture probability for $\pi^-$ at rest in the target is extremely high, the neutrino spectrum from the stopped pions will be dominated by the decay of $\pi^+$ by a factor of about $10^4$.
Figure \ref{pionspectrum} shows the neutrino spectra from the $\pi^+$ $\rightarrow$ $\mu^+$ $\rightarrow$ $e^+$ decay.
  Noteworthy in this decay is the absence of electron antineutrinos, making this 
  source a favored means of looking for appearance of \nuebar, again utilizing 
  IBD in a suitable neutrino detector.  

  These neutrino sources are isotropic, there is no kinematic directionality to define a Òbeam.Ó  
  As a result, the efficiency of detection is directly related to the solid angle subtended by 
  the detector, placing high emphasis on having the source as close to the detector as possible.  
  In the case of IsoDAR this distance is a few meters from the detector surface 
  (16.5 meters from the center of the KamLAND fiducial volume), in the case of 
  \DD the baseline is 20 km from the large water-Cherenkov counter (assumed to be Hyper-K). 
   As the principal goals of these experiments is oscillation physics, the driving term is {\em L/E}, 
   the baseline distance divided by the neutrino energy.  If {\em E} is low, the baseline {\em L} can also 
   be low to preserve the same ratio.  As a consequence, the 20 km baseline and 45 MeV 
   average \numubar energy addresses the same oscillation point as the 1300 km, 
   3 GeV DUNE beam, or the 300 km, ~500 MeV T2K beam.

The premise of these experiments is that relatively small and compact sources of neutrinos 
can be built and installed at the proper distances from existing or planned large water- or
 liquid-scintillator-based
neutrino detectors, providing access to the physics measurements with substantially 
reduced costs.  With respect to the long-baseline experiments (e.g. T2K) the 
beamlines from the major accelerator centers operate much more efficiently and cleanly 
in the neutrino mode, while the DAR measurements, utilizing IBD, address only the
 anti-neutrino mode.  Consequently, installing \DD cyclotrons at the proper distance 
 from the long-baseline detectors, and operating the neutrino beams simultaneously, 
 offers a huge improvement in the sensitivity and data rates over the individual experiments.  
 Discrimination of the source of events is straightforward, both from the energy deposition 
 of events from each source, as well as from timing:  neutrinos from the cyclotrons are 
 essentially continuous (up to 100$\%$ duty factor), while those from the large accelerators 
 are tightly pulsed with a very low overall duty factor.

  Nevertheless, the lack of directionality of DAR neutrinos, and the small solid angle 
  between source and detector calls for the highest-possible flux from the source to 
  ensure meaningful data rates.  Available accelerator technologies and design 
  configurations have been explored, for beam current performance, 
  cost and footprint; we have arrived at the choice of compact \mbox{cyclotrons\hspace{2pt}\cite{costComparison}}.  
  The only deficiency of this option is the average current.  For appropriate data rates, 
  our specification is 10 mA of protons on target.  This pushes the highest current from 
  cyclotrons by about a factor of 3,\footnote{Isotope-producing \hminus cyclotrons rarely reach
  2 mA, the current record-holder for cyclotron current 
  is the \mbox{3 mA} PSI Injector 2, a 72 MeV separated-sector proton cyclotron 
  injecting the 590 MeV Ring Cyclotron.}
   and much of the accelerator development work of our 
  group to date has been devoted to addressing the factors that limit the maximum current 
  in compact \mbox{cyclotrons\hspace{3pt}\cite{BCS}$^,\hspace{1pt}$\cite{RFQ}$^,\hspace{1pt}$\cite{simulations}}.

   \begin{wrapfigure}{r}{0.5\textwidth}
  \begin{center}
  \vspace*{-10mm}
  \includegraphics[width=0.5\textwidth]{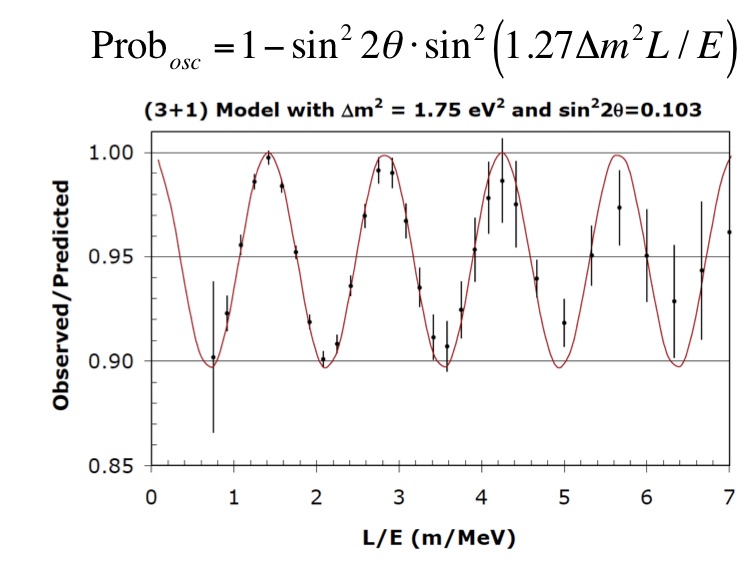}
  \end{center}
  \vspace*{-6mm}
  \caption{ Oscillations seen in KamLAND for a 5 year IsoDAR run, for the global fit parameters still consistent with the IceCube analysis.  IBD event rate is about 500 per day.}
    \label{IsoDARosc}
    \vspace*{-2mm}
  \end{wrapfigure}


 In the next section the physics rationale for the IsoDAR and \DD experiments 
  will be briefly described, while subsequent sections will address the configuration 
  of the cyclotrons, and progress made in pushing the current 
  limits from cyclotrons to the required level.  The IsoDAR target will be described, 
   capable of handling the 600 kW of proton beams and optimized for $^8$Li production.  
  Finally, the IsoDAR experiment will be compared with other ongoing initiatives for 
  searching for sterile neutrinos.

  \begin{wrapfigure}{r}{0.6\textwidth}
  \begin{center}
  \vspace*{-4mm}
  \includegraphics[width=0.6\textwidth]{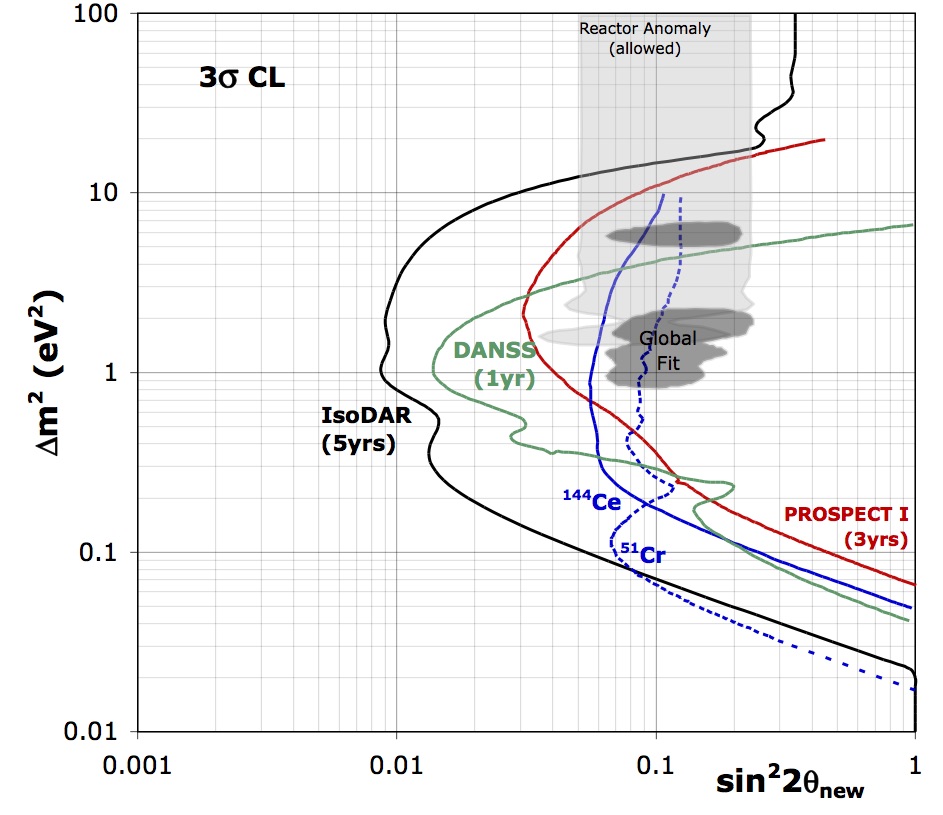}
  \end{center}
  \vspace*{-6mm}
  \caption{ Sensitivity of 5 year IsoDAR run compared to other sterile neutrino experiments.  
  DANSS is a reactor experiment in Kalinin (Russia)\hspace{2pt}\protect\cite{DANSS}; $^{144}$Ce and $^{51}$Cr are the SOX 
  experiment at Borexino (Gran Sasso, Italy)\hspace{1pt}\protect\cite{SOX}, 
  PROSPECT is a reactor experiment at HFIR at ORNL (USA)\hspace{1pt}\protect\cite{Prospect}.
     \label{IsoSens}}
    \vspace*{-6mm}
  \end{wrapfigure}

\section{Neutrino Measurements}
  
\subsection{IsoDAR}

Anomalies in \nuebar disappearance rates have been observed in reactor and radioactive 
source \mbox{experiments\hspace{2pt}\cite{reactorAnomaly}}.  Postulated to explain these has been
the existence of one or 
more sterile neutrinos, that do not in themselves interact in the same manner as ``active''
neutrinos (hence are called 
``sterile''), however the active neutrinos can oscillate through these 
sterile states, and in this manner affect the ratio of appearance and disappearance from 
the known three flavor eigenstates.  \mbox{Global fits\hspace{2pt}\cite{globalFit}} of data from experiments point to a mass
 splitting in the order of 1 to almost 8 {\em eV\hspace{2pt}$^2$}, and a {\em sin$^2$(2\hspace{2pt}$\theta$)} of 0.1. 
  Recent analysis of IceCube \mbox{data\hspace{2pt}\cite{IceCube}}, exploiting a predicted resonance in the MSW matrix for
   \numubar passing through the core of the earth appear to rule out {\em $\Delta$m$^2$} values of 
   1 {\em eV\hspace{2pt}$^2$} or below, however values above this energy are still possible.  
   
  The very large {\em  $\Delta$m$^2$} imply a very short wavelength for the oscillations, in fact for the $^8$Li
 neutrino it is measured in meters, so within the fiducial volume of KamLAND one could see 
 several full oscillations.  Folding in the spatial and energy resolutions of the KamLAND 
 detector ({\em 12~cm/$\sqrt{E_{MeV}}$}) and ({\em 6.4\%/$\sqrt{E_{MeV}}$})
  respectively, the expected neutrino interaction pattern for the case of $\Delta$m$^2$ = 1.75 {\em eV\hspace{2pt}$^2$} is shown in Figure \ref{IsoDARosc}.

  Figure \ref{IsoSens} shows a sensitivity plot for 
  IsoDAR, this experiment covers very well the regions of interest for sterile neutrinos.

\subsection{Layout of \DD Experiment}

Search for CP violation in the lepton sector has been a high priority for many years. 
 \DD combined with a long-baseline beam (e.g. T2K @ Hyper-K operating in neutrino mode only) 
 can in 10 years cover almost all of the $\delta$ CP-violating phase angles.\cite{discoveryPlot}

 \begin{wrapfigure}{r}{0.4\textwidth}
  \begin{center}
  \vspace*{-8mm}
  \includegraphics[width=0.4\textwidth]{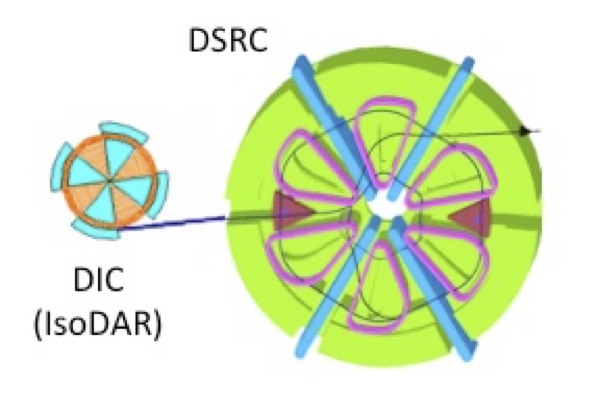}
  \end{center}
  \vspace*{-6mm}
  \caption{Schematic of the two cyclotrons in a \DD module.  The injector (DIC - \DD Injector Cyclotron) also serves as the proton source for IsoDAR.  The DSRC (\DD Superconducting Ring Cyclotron) produces protons at 800 MeV.
     \label{fig:cycl}}
    \vspace*{-1mm}
  \end{wrapfigure}

 The experimental configuration includes three stations, each with identical targets that provide 
 neutrino sources (from stopped $\pi^+$), one at 1.5 km (essentially as close to the 
 detector as feasible) that normalizes the flux seen in the detector, one at 8 km that catches 
 the rise in the \nuebar appearance, and the principal station at 20 km, which measures 
 the \nuebar appearance at the peak of the oscillation curve.  The absolute appearance 
 amplitude is modulated by the CP-violating phase.  
 The current on target, hence the neutrino flux, 
 is adjusted sequentially at each station (by ``beam-on'' timing) to be approximately equivalent to the flux 
 from the long-baseline beam. The total timing cycle from all stations allows approximately 
 40\% of time when none are delivering neutrinos, for background measurements.


\section{Cyclotron Configuration}

Figure \ref{fig:cycl} shows schematically the basic configuration of a cyclotron ``module'' for \DD,
 showing the ``chain'' of injector-booster cyclotron with a top energy of 60 MeV, 
 and the main \DD superconducting ring cyclotron (DSRC) which delivers 800 MeV protons to
  the pion-production target.  Note that the injector cyclotron is exactly the machine 
  that is needed for the IsoDAR experiment, so developing this cyclotron is a direct step in the path towards \DD.
  
    \begin{table}[h]
\caption[]{The most relevant parameters for the IsoDAR and \DD cyclotrons.  
IsoDAR has a single station with one cyclotron, \DD has three stations, at 1.5, 8, and 20 km
from the detector.  The first two stations have a single cyclotron pair (DIC and DSRC), the 20 km station
has two cyclotron pairs for higher power. Though the total power is high, because the targets are large and the beam is uniformly spread over the target face, the power {\em density} is low enough to be handled by conventional engineering designs. The \DD target has a long conical reentrant hole providing a very large surface area.}
\label{tab:CyclTable}
\vspace{0.4cm}
\begin{center}
\begin{tabular}{|l|c|c|l|}
\hline
& &  \\
&
IsoDAR &
\DD 
\\ \hline
Particle accelerated &
\htp & \htp \\
Maximum energy & 60 MeV/amu & 800 MeV/amu \\
Extraction & Septum & Stripping \\
Peak beam current (\htp) & 5 mA & 5 mA \\
Peak beam current (proton) & 10 mA & 10 mA \\
Number of stations & 1 & 3 \\
Duty factor & 100\% & 
\begin{minipage}{2.5in}\setstretch{0.8}\begin{center}
15\% - 50\% \\
(time switching between 3 stations)
\end{center}
\end{minipage} \\
Peak beam power on target & 600 kW & 8 MW \\
Peak power density on target & 2 kW/cm$^2$  & $\approx$ 2 kW/cm$^2$ \\
Average beam power on target & 600 kW & 1.2 to 4 MW \\
Maximum steel diameter & 6.2 meters & 14.5 meters \\
Approximate weight & 450 tons & 5000 tons \\ \hline
\end{tabular}
\end{center}
\end{table}

  Table \ref{tab:CyclTable} lists high-level parameters for the IsoDAR and \DD cyclotrons.  
  Note the power implication of delivering 10 mA to the production targets.  
  These very high power-requirements call for minimizing
beam loss during the acceleration and transport process.  
Any beam loss is not only destructive of components, 
but also activates materials and greatly complicates maintenance of accelerator systems.
Some beam loss is unavoidable, however by appropriate use of cooled collimators and beam dumps,
 and by restricting as much as possible these losses to the lower energy regions of the cyclotrons, 
 the thermal and activation damage can be minimized.

The single biggest innovation in these cyclotrons, aimed at increasing the maximum current,
is the use of \htp ions\hspace{2pt}\cite{LucianoHtp} instead of protons or \hminus. 
 As the biggest source of beam loss is space charge blowup at low energies,
  the lower {\em q/A} (2 protons for a single charge), and higher mass per ion
   (= 2 amu - atomic mass units) greatly reduces the effects of the repulsive forces of the 
   very high charge in a single bunch of accelerated beam. This helps keep the size of the
    accelerated bunches down so there will be less beam lost on the inside of the cyclotron. 
     Keeping the molecular ion to the full energy also allows for stripping extraction at 
     800 MeV/amu, reducing beam loss in the extraction channels.
     
     While the size and weight of these cyclotrons may appear large, there are 
     examples of machines of comparable size that can serve as engineering models for beam dynamics,
     magnetic field design and costing.  The PSI Injector 2, a 72-MeV 3-mA machine models some
     aspects of the IsoDAR cyclotron relating to the RF system and space-charge dominated
     beam dynamics\hspace{2pt}\cite{AdelmanInjII}.  Magnet design and steel size/weight bear some 
     similarities to IBA's 235 MeV proton radiotherapy cyclotron\hspace{2pt}\cite{IBA235}.  
     The DSRC bears significant similarities to the superconducting ring cyclotron 
     at RIKEN\hspace{2pt}\cite{RIKENSRC}.  While this cyclotron is designed for uranium beams, 
     so the beam dynamics are not directly relevant, the cryostat and magnet designs 
     are extremely close to the \DD requirements, and so serve as a good engineering 
     and costing model for the DSRC.  
    
\section{IsoDAR developments}

As indicated above, efforts of our group have focused on producing high currents of \htp
for injection into the IsoDAR cyclotron, modeling the capture and acceleration of these ions,
and on the design of the target for handling 600 kW of proton beam and maximizing the 
production of $^8$Li to generate the \nuebar flux delivered to KamLAND.
   
\subsection{Producing High Currents of H$_2^+$\xspace for Injection}

Experiments at the Best Cyclotron Systems, Inc. test stand in Vancouver, BC \cite{BCS} tested the 
VIS high-current proton source\hspace{2pt}\cite{VIS} for its performance in generating \htp  beams.  
Our requirement for \htp is a maximum of 50 mA of continuous beam from the source, 
which would provide an adequate cushion in the event that capture into the cyclotron 
cannot be enhanced by efficient time-bunching of the beam (see next section).  
The VIS only produced about 15 mA of \htp (while we did measure 40 mA of protons); 
using this source would require efficient bunching.  To increase our safety margin, 
a new ion source, labeled ``MIST-1'' has been built\hspace{2pt}\cite{MISTI} based on an
 LBL-developed filament-driven, multicusp design\hspace{2pt}\cite{Leung} which demonstrated a 
 much more favorable p/\htp ratio, and currents in the range required.  
 This source has been designed with a high degree of flexibility, 
 to adjust geometric, magnetic field and plasma conditions to optimize \htp performance.  
 It is now being commissioned.

\subsection{Capturing and Accelerating High Currents of H$_2^+$\xspace}

\begin{wrapfigure}{r}{0.6\textwidth}
  \begin{center}
  \vspace*{-4mm}
  \includegraphics[width=0.6\textwidth]{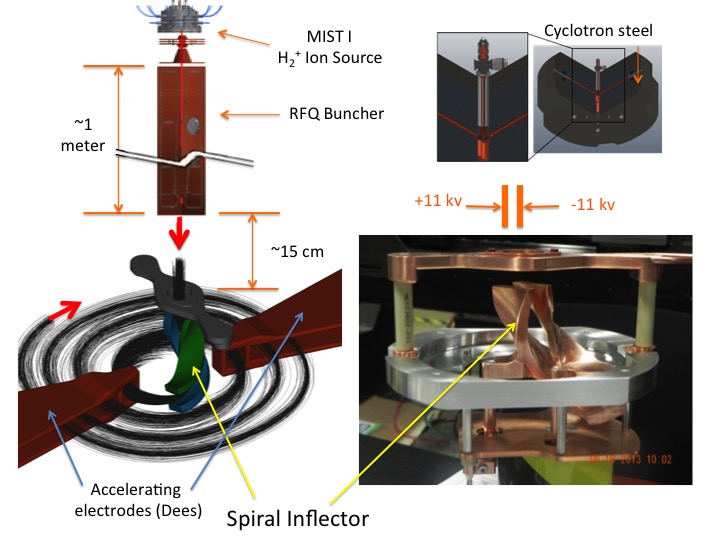}
  \end{center}
  \vspace*{-6mm}
  \caption{Low energy injection line and central region of the DIC.  A short transport line connects the MIST-1 \htp ion source with the RFQ buncher, which compresses the beam into packets of 
  about \mbox{$\pm$ $15^{\circ}$}.  These packets are fed to the spiral inflector (photographed in lower-right), electrostatic deflector plates that bend the beam into the plane of the cyclotron.  The distance from the end of the RFQ to the accelerating dees must be kept to a minium as there is energy spread in the beam and long transport distances will cause the beam to debunch.  As a result the RFQ must be installed largely inside the steel of the cyclotron (pictured in upper right).
     \label{fig:CentralReg}}
    \vspace*{-3mm}
  \end{wrapfigure}

Cyclotrons accelerate beam via RF (radio-frequency, for our cyclotron around 50 MHz) fields applied to electrodes (called ``Dees'') extending along the full radial extent of the beam.  Particles reaching the accelerating gap at the right phase of the RF will receive a positive kick, while those arriving outside this phase angle will be decelerated and lost.  The phase acceptance of the cyclotron is typically about $\pm$ $15^{\circ}$, so if the injected beam is not bunched longitudinally, only 10\% of a continuous beam will be accepted.  Hence the need for 50 mA of unbunched beam.  Bunching is conventionally done with a double-gap RF cavity placed about one meter ahead of the injection point.  Maximum efficiency improvement is no more than a factor of 2 or 3.  

A novel bunching technique using an RFQ was proposed many years ago\hspace{2pt}\cite{Hamm} that could in principle improve bunching efficiency to almost 85\%.  We have recently been awarded funding from NSF to develop this technique, and are working with the original proponent, and other key RFQ groups in the US and Europe to build and test this new buncher.  Figure \ref{fig:CentralReg} shows schematically the central region of the cyclotron, including the MIST-1 source, the RFQ, and spiral inflector that bunches and bends the beam into the plane of the cyclotron.

Once inflected into the plane of the cyclotron, the beam must be stably captured and accelerated to the full energy and extraction radius (of 2 meters in our case).  In addition, there must be adequate turn separation at the outer radius to cleanly extract the beam.  The particles experience 96 turns from injection to extraction, and the radial size of the beam must be controlled so that a thin septum can be inserted between the 95th and 96th turns that will not intercept any appreciable amount of beam.  With a total of 600 kW, even a fraction of a percent of beam lost on this septum can damage it.

\begin{wrapfigure}{r}{0.4\textwidth}
  \begin{center}
  \vspace*{-10mm}
  \includegraphics[width=0.4\textwidth]{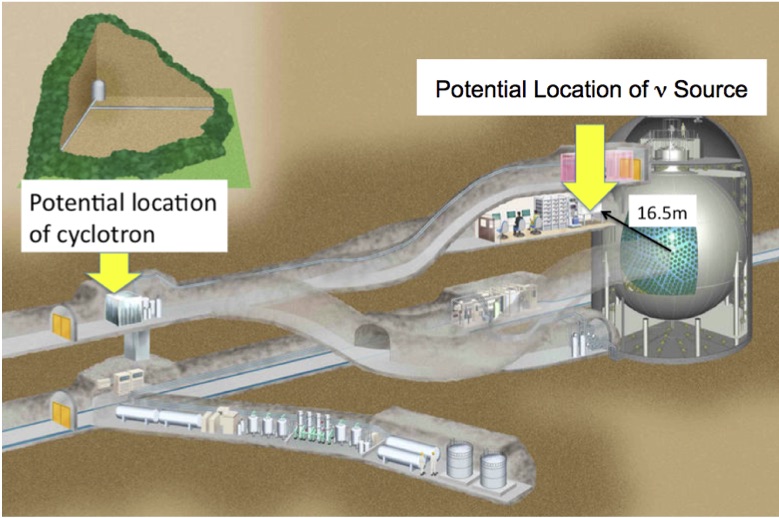}
  \end{center}
  \vspace*{-4mm}
  \caption{Configuration of IsoDAR on the KamLAND site.  
  \label{fig:layout}}
    \vspace*{-4mm}
  \end{wrapfigure}

Extensive simulations, using the OPAL code\hspace{2pt}\cite{OPAL} developed at PSI specifically for beam-dynamics of highly space-charge-dominated beams in cyclotrons have been used to show that this is possible, and to locate collimators and scrapers in the first few turns to control beam halo (that would be intercepted on the extraction septum).  This code has also shown that space-charge forces can actually contribute to stability of the accelerating bunch by introducing a vortex motion within the bunch that limits longitudinal and transverse growth of the bunch\hspace{2pt}\cite{Jacob}.

These developments  give us confidence that the technical specifications for the IsoDAR cyclotron can be met.

\subsection{Target design}

The configuration of the IsoDAR experiment is shown in Fig \ref{fig:layout}.  
The cyclotron is located in a vault previously used for water purification, 
the target is located in one of the construction drifts repurposed as a control room that is no longer used.

\begin{wrapfigure}{l}{0.6\textwidth}
  \begin{center}
  \vspace*{-3mm}
  \includegraphics[width=0.6\textwidth]{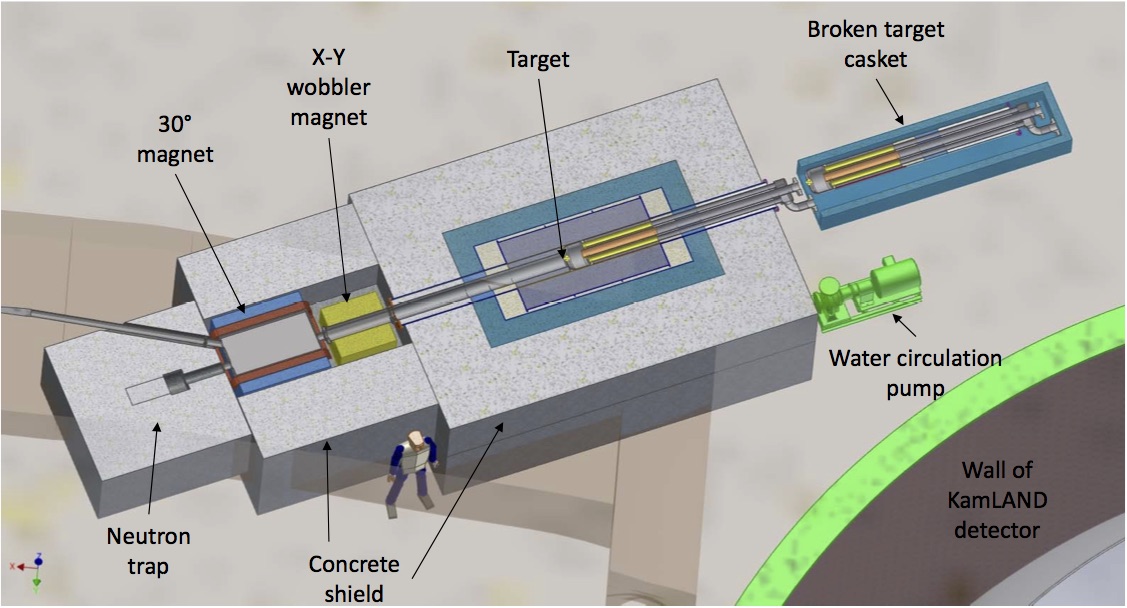}
  \end{center}
  \vspace*{-4mm}
  \caption{Target/sleeve/shielding structure.  The target is 16.5 meters from the center of the KamLAND fiducial volume.  Beam is bent 30$^{\circ}$ to the target providing shielding for backstreaming neutrons. A wobbler magnet spreads beam out on the 20 cm diameter target face. The target assembly can be pulled from the back of the structure into a casket. This hole is also shielded with removable concrete blocks.  The shielding structure consists of steel and borated concrete.
  \label{fig:Tgt@KamLand}}
    \vspace*{-4mm}
  \end{wrapfigure}

Beam is extracted from the cyclotron and transported about 50 meters to the target located 
close to the KamLAND detector.  The 5 mA of \htp is stripped in this transport line, 
the resulting 10 mA of protons are directed to the beryllium target.  
Beryllium is a very efficient neutron producer, for the 60 MeV proton beam the yield is 
approximately 1 neutron per 10 protons.  
These neutrons stream through to the sleeve surrounding the target, 
containing small beryllium spheres (less than 1 cm diameter) surrounded by 
highly-enriched $^7$Li (99.995\%) .  
The sleeve is a cylinder 50 cm in radius and 2 meters long, 
and is surrounded by a 5 cm graphite reflector.  
Shielding outside the reflector consisting of iron and borated concrete which 
contains the neutron flux to limit neutrons reaching the rock walls.

Fig \ref{fig:Tgt@KamLand} shows the target, sleeve and shielding assembly in relation 
to the KamLAND detector.  
The $^8$Li yield from the moderated and captured neutrons varies with the 
fractional composition of beryllium and lithium in the sleeve, 
the maximum is about 3\% ($^8$Li per incident proton on target) for 30\% (by weight) of lithium.  
This is close to the interstitial volume of tightly packed spheres. 
All numbers are based on GEANT4 calculations\hspace{2pt}\cite{Adriana}.

\begin{wrapfigure}{r}{0.4\textwidth}
  \begin{center}
  \vspace*{-8mm}
  \includegraphics[width=0.4\textwidth]{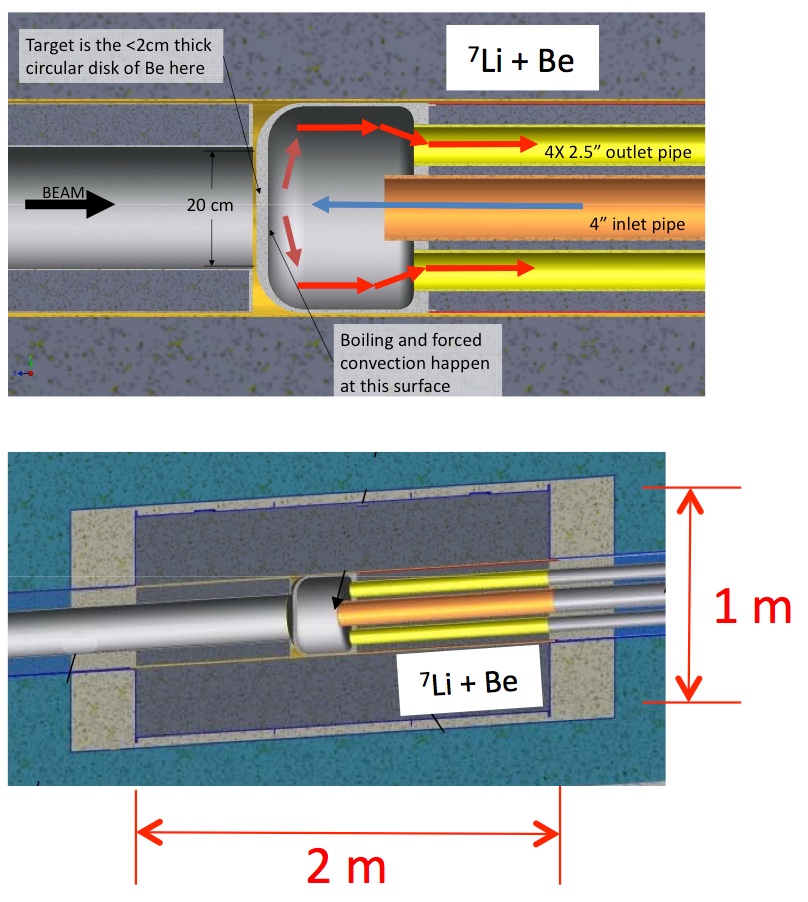}
  \end{center}
  \vspace*{-4mm}
  \caption{Section through target and sleeve.
  \label{fig:tgt}}
    \vspace*{-15mm}
  \end{wrapfigure}

Fig \ref{fig:tgt} shows the target assembly, a spun-cast beryllium piece with the front surface 
(where the beam hits) being 1.8 cm thick (range of protons is 2 cm, 
so Bragg peak, at energy too low to efficiently produce neutrons, is in the cooling water, 
reducing heat load in target.  A jet of heavy water is directed to the back surface
 of the target in a manner that effectively removes the 600 kW of beam power to a heat exchanger. 
  The thermal behavior of the target is being modeled and will be experimentally tested in the future.

\section{IsoDAR Compared with other Sterile Neutrino Experiments}

Table \ref{tab:Compare} compares the IsoDAR experiment with two other sterile-neutrino search experiments, SOX\hspace{2pt}\cite{SOX} and DANSS\hspace{3pt}\cite{DANSS}.  Sensitivity comparisons were given in Figure \ref{IsoSens}, the table highlights some of the rationale for the significantly higher sensitivity of IsoDAR.

    \begin{table}[h]
\caption[]{Comparison of IsoDAR with SOX, the $^{144}\mathrm{Ce}$  experiment at Borexino, and DANSS, a representative reactor experiment. Relative sensitivities of these three experiments were shown in Fig. \ref{IsoSens}}.
\label{tab:Compare}
\vspace{0.2cm}
\begin{center}
\begin{tabular}{|l|c|c|c|}

\hline
& & &  \\
&
\textbf{IsoDAR} &
\textbf{SOX} & \textbf{DANSS}
\\ \hline
 & & & \\
\textbf{SOURCE} &
$^8\mathrm{Li}$ & $^{144}\mathrm{Ce}$ & Fuel burning \\ 
 & & & \\
Spectral purity & Clean $\beta$ spectrum & Clean $\beta$ spectrum & complex, with anomalies \\
 & & & \\
Rate stability & \begin{minipage}{1.2in}\setstretch{0.8}\begin{center} Stable, dependent\\on accelerator \end{center}\end{minipage} & 
\begin{minipage}{1.2in} \setstretch{0.8}\begin{center} Decays with \\ 285 day halflife  \end{center}\end{minipage} & 
\begin{minipage}{1.1in}\setstretch{0.8}\begin{center} Changes with \\ fuel aging \end{center} \end{minipage} \\
 & & & \\
\begin{minipage}{1in} \setstretch{0.8}\begin{center} Energy of \nuebar \\ flux maximum \end{center}\end{minipage} 
& 8.5 MeV & 3.4 MeV & 3.5 MeV  \\ 
 & & & \\
\textbf{DETECTOR} & KamLAND  & Borexino & Solid scintillator \\ 
 & & & \\
Volume & 900 tons & 100 tons & $<$10 tons \\ 
 & & & \\
Neutron bkgnd & \begin{minipage}{1.4in}\setstretch{0.8}\begin{center} Manageable \\ shield design \end{center}\end{minipage} &  
\begin{minipage}{1.4in}\setstretch{0.8}\begin{center} Manageable \\ shield design \end{center}\end{minipage} &
 \begin{minipage}{1.6in}\setstretch{0.8}\begin{center} Difficult to shield, limits\\proximity to core \end{center}\end{minipage} \\ 
  & & & \\
 \begin{minipage}{1.2in}\setstretch{0.8}\begin{center} Cosmic bkgnd \\ (rock overburden)\end{center}\end{minipage} & 2700 MWE & 3400 MWE & \begin{minipage}{1.5in}\setstretch{0.8}\begin{center} shallow, \\ high muon rates \end{center}\end{minipage} \\ \hline


\end{tabular}
\end{center}
\end{table}

In summary, IsoDAR is a very compelling experiment for the search for sterile neutrinos, but because of the high event rates and excellent statistics, the reach of physics for this extremely short baseline configuration extends to non-standard interactions, spectral shape and other neutrino-characterization experiments as well.  The challenging technologies for producing the high-power beams and optimizing neutrino production are being developed at a steady pace, ever increasing the feasibility of these experiments.

\section*{Acknowledgments}

Work supported by the US National Science Foundation under Grant No. NSF-PHY-1505858, and by the MIT Bose Foundation.

\end{document}